\documentclass[reprint,twocolumn
 amsmath,amssymb,
 aps,
]{revtex4-2}
\usepackage[compact]{titlesec}
\titleformat{\section}
  {\normalfont\normalsize\bfseries\uppercase}{\thesection.}{1em}{}
\titleformat{\subsection}
  {\normalfont\normalsize\bfseries}{\thesubsection.}{1em}{}
\titleformat{\subsubsection}
  {\normalfont\normalsize\itshape}{\thesubsubsection.}{1em}{}
\usepackage{dcolumn}
\usepackage{bm}
\usepackage{xcolor}
\usepackage{hyperref}
\usepackage{mathrsfs}
\usepackage{mathtools}

\begin{document}


\title{\textbf{The Klein-Gordon equation with relativistic mass: a relativistic Schr\"odinger equation} }

\author{P.-A. Gourdain}
 \email{Contact author: gourdain@pas.rochester.edu}
\affiliation{%
 Physics and Astronomy Department,\\
 Laboratory for Laser Energetics,\\
 University of Rochester, New York 14627, USA
}%

\date{\today}

\begin{abstract}
The Klein-Gordon equation describes the wave-like behavior of spinless particles since it is Lorentz invariant. While it seemed initially ripe for explaining the electronic structure of the hydrogen atom, the lack of a unconditional positive probability density really limited its applications. Yet, it is intimately connected with fermions. Any solution to the Dirac equation is automatically a solution to the Klein-Gordon equation. What is even more surprising, the Klein-Gordon equation for a free particle turns into the Schr\"odinger equation in the non-relativistic limit. In this work we show that these problems disappear when we use the relativistic mass instead of the rest mass. While the Klein-Gordon equation losses its Lorentz invariance because of this transformation, it gains most of the features present the Schr\"odinger equation, including the unconditional positivity of probability density, while keeping most of its relativistic characteristics intact, including the matter-wave dispersion relation. 
What is even more surprising, the non-relativistic, quasi-static limit of the Klein-Gordon equation with relativistic mass is simply the Schr\"odinger equation under all possible conditions. 
So, it can be argued that this Klein-Gordon equation is a sort of relativistic Schr\"odinger equation.
\end{abstract}

\maketitle
\section*{Introduction}
The Klein-Gordon equation \cite{gordon1926,klein1927} was initially seen by Schr\"odinger as a possible means to explain the electronic structure of the hydrogen atom. While Schr\"odinger had discovered this equation before Klein and Gordon, the lack of agreement between theory and experiment led him to devise his namesake equation \cite{schrodinger1926} instead. It was found later that the Klein-Gordon equation actually describes spinless particles since it is Lorentz invariant\cite{jordan1964}. 

Yet, any solution to the Dirac equation \cite{dirac} is also a solution to the Klein-Gordon equation. Further, its dispersion relation in vacuum matches the matter-wave theory of de Broglie \cite{debroglie1924recherches}. It is also recognized as a direct relativistic generalization of the free-particle Schr\"odinger equation \cite{strange1998relativistic}. Finally, it has a strong connection with condensed matter physics \cite{caudrey1975sine} and lattice dynamics \cite{dodd1982solitons}. It is possible to solve this equation when using different type of potentials \cite{alhaidari2006dirac,saad2008klein,onate2016analytical}, or adding nonlinearities \cite{dehghan2009numerical} or time fractional derivatives \cite{golmankhaneh2011nonlinear,partohaghighi2022fractal} using new numerical algorithms \cite{bratsos2009numerical,golmankhaneh2011nonlinear,khader2014accurate}.

As a result, this equation should be at the heart of quantum mechanics. However, it is not the case, mostly because the density $\rho$ is not always positive and cannot be interpreted as an actual probability density, a cornerstone of the Schr\"odinger equation. In this paper, we show that this major problem disappears as soon as the relativistic mass is used instead of the rest mass. This simple transformation allows the density to remain positive and this density can now be interpreted as a probability density. The only requirement for this  probability density to be conserved is energy conservation.

After this introduction, the paper lays the foundations necessary to construct the Klein-Gordon equation with relativistic mass, then derives its continuity equation and verifies that the density is conserved and always positive. Finally, a parallel with electromagnetism highlights a key feature behind the space-time coupling.


\section*{The Klein-Gordon equation with relativistic mass}
We start with the total relativistic energy $E$,
\begin{equation}\label{eq:relativistic_energy}
    E^2=\textbf p^2c^2+m_0^2c^4,
\end{equation}
 of a particle with charge $q$, a mass at rest $m_0$, and moving with a velocity $\textbf v$. Here $\textbf p=m\textbf v$ is the relativistic momentum and $m$ the relativistic mass, i.e. $m=\gamma m_0$, where $\gamma$ is the Lorentz factor. We want to remove the rest mass from both sides of this equation to get a quantum equivalent equation where only the relativistic mass is present. To do so, we need to isolate the rest mass from the total relativistic energy, and then define all quantities as a function of the relativistic mass $m$.
\subsection*{Isolation of the rest mass inside the relativistic energy}
Using the relativistic Lagrangian $L_R$ defined by
\begin{equation}\label{eq:relativistic_Lagrangian}
  L_R=-\frac{m_0c^2}{\gamma},  
\end{equation}
it can be shown that the total relativistic energy of Eq. \eqref{eq:relativistic_energy} is a Hamiltonian of motion when energy is conserved, since $E=\partial_{\dot {\textbf r}} L_R\cdot\dot {\textbf r}-L_R$. Further, since any Lagrangian $L_R$ is indefinite with respect to addition of a constant kinetic energy \cite{cline2021}, we can define a new energy $U$ simply by adding $m_0c^2$ to $L_R$,
i.e. $U=\partial_{\dot {\textbf r}} (L_R+m_0c^2)\cdot\dot {\textbf r}-(L_R+m_0c^2) \text{ or } U=E-m_0c^2$.
Using this new energy we get \cite{greiner2000relativistic}
\begin{equation}\label{eq:relativistic_energy_using_H}
    \left(m_0c^2+U\right)^2=\textbf p^2c^2+m_0^2c^4.
\end{equation}

\subsection*{Mass for a charged particle inside an electromagnetic potential}
When a particle with rest mass $m_0$ and charge $q$ is moving inside an electromagnetic field with scalar potential $\phi$ and vector potential $\textbf A$, the Lagrangian $L_P$ of the particle is given by \cite{Hand1998}, 
\begin{equation}\label{eq:lagrangian}
L_P=L_R-q\phi+q\textbf A\cdot\dot {\textbf r},
\end{equation}
When the potentials $\phi$ and $\textbf A$ depend explicitly neither on time nor particle velocities, the total energy of the particle,
\begin{equation*}
    \mathscr E_P=\partial_{\dot {\textbf r}} L_P\cdot\dot {\textbf r}-L_P=mc^2+q\phi,
\end{equation*}
is conserved. First, we suppose the particle to be at rest and located at infinity, where $\phi_0=0$. In this case, the total energy of the particle $\mathscr E_P$ is 
\begin{equation*}
    \mathscr E_P=m_0c^2
\end{equation*}
Now, for this particle to move inside the potentials $\phi$ and $\textbf A$, we need $q\phi<0$. This requirement can be deduced explicitly from the fact that $m\geqslant m_0$ since $\gamma\geqslant 1$. In this case
\begin{equation*}
    \mathscr E_P=mc^2+q\phi
\end{equation*}
Because the energy of the particle is conserved, we can link the initial (at infinity) and final states, yielding 
\begin{equation}\label{eq:energy_conservation}
    m_0c^2=mc^2+q\phi.
\end{equation}

Now, we may want to work with an electromagnetic field that varies in time. In this case, this is not the particle energy $\mathscr E_P$ is not conserved but this is the total energy $\mathscr E_T$ of the particle at rest \textit{and} the electromagnetic field that is conserved. As before, we first compute the total energy at infinity 
\begin{align*}
    \mathscr E_T=m_0c^2+\frac{1}{2}\int\left[\varepsilon_0\textbf E_0^2+\frac{\textbf B_0^2}{\mu_0}\right]dV.
\end{align*}
Here $\textbf E_0(t,x,y,z)$, $\textbf B_0(t,x,y,z)$ are the total electric and magnetic fields with the particle at infinity. When the particle moves inside $\phi$ and $\textbf A$ we have
\begin{align*}
    \mathscr E_T=mc^2+q\phi+\frac{1}{2}\int\left[\varepsilon_0\textbf E^2+\frac{\textbf B^2}{\mu_0}\right]dV,
\end{align*}
where $\textbf E(t,x,y,z)$ and $\textbf B(t,x,y,z)$ are the total electric and magnetic fields when the particle is moving inside the potentials $\phi$ and $\textbf A$. Note that the electric and magnetic fields of the particle are included in all these electric and magnetic fields. We now define an effective mass 
\begin{equation}\label{eq:effective_mass}
    \tilde m=m+\frac{1}{2c^2}\int\left[\varepsilon_0\left(\textbf E^2-\textbf E_0^2\right)+\frac{\textbf B^2-\textbf B_0^2}{\mu_0}\right]dV,
\end{equation}
which encapsulates these non-conservative processes. If we suppose that there is no outgoing energy flux at infinity (full system isolation hypothesis), the Poynting theorem guarantees the conservation of $\mathscr E_T$, and we have 
\begin{equation}\label{eq:non_conservative_energy}
    m_0c^2=\tilde mc^2+q\phi.
\end{equation}
Despite the existence of non conservative effects, it is interesting to note that the equation above yields
\begin{equation}\label{eq:mass_potential_connection}
    \partial_t\tilde m=-\frac{q}{c^2}\partial_t\phi.
\end{equation}
We can turn any equation with non-conservative effects to an equation with conservative effects simply by setting $\tilde m\rightarrow m$. However, conservative effects also require $\phi$ and $\textbf A$ to be time independent. To avoid any inconsistencies between time dependent and time-independent gauges, we chose to use the Lorenz gauge \cite{lorenz}  
\begin{equation}\label{eq:Lorenz}
    c^{-2}\partial_t\phi+\nabla\cdot\textbf A=0
\end{equation}
throughout, which turns into the Coulomb gauge 
\begin{equation*}
    \nabla\cdot\textbf A=0
\end{equation*}
for steady state fields. Further, if we use Eqs. \eqref{eq:mass_potential_connection} and \eqref{eq:Lorenz} we get
\begin{equation}\label{eq:mass_vector_potential_connection}
    \partial_t\tilde m=q\nabla\cdot\textbf A,
\end{equation}
which again is consistent when we transition to time independent phenomena. 
Further, we can obtain a non-relativistic equation by setting $\tilde m\rightarrow m_0$, and a quasi static version by setting $c\rightarrow +\infty$. 


\subsection*{Derivation of the Klein-Gordon equation with relativistic mass}
Using minimal coupling, we can replace the particle energy with the quantum energy operator (i.e. $U\rightarrow \hat U-q\phi$), its momentum with the momentum operator (i.e. $\textbf p\rightarrow\hat{\textbf p}-q\textbf A$), and its rest mass with the effective mass from Eq. \eqref{eq:non_conservative_energy} inside Eq. \eqref{eq:relativistic_energy_using_H}, and we get 
\begin{equation}\label{eq:compact_KG}
    \left[\left(\tilde mc^2+\hat U\right)^2-\left(\hat{\textbf p}-q\textbf A\right)^2c^2\right]\psi=\left[\tilde mc^2+q\phi\right]^2\psi.
\end{equation}
Here the quantum energy operator $\hat U$ is $i\hbar\partial_t$, $\hat{\textbf p}$ is the quantum momentum operator $\hat{\textbf p}=-i\hbar\nabla$, and $\psi$ is the function that capture the wave behavior of the particle. Since the mass depends on time explicitly, $\hat U$ and $\tilde m$ do not commute and we must rewrite Eq. \eqref{eq:compact_KG} as
\begin{equation}\label{eq:expanded_KG}
\begin{split}
        \left[i\tilde m\hbar\partial_t-\frac{\hbar^2}{2c^2}\partial_{tt}-\frac{1}2(\hat{\textbf p}-q\textbf A)^2\right]\psi=\\
    \left[\tilde mq\phi+\frac{1}2\left(\frac{q^2\phi^2}{c^2}-i\hbar\partial_t\tilde m\right)\right]\psi.
\end{split}
\end{equation}
We can expand the momentum operator as 
\begin{equation*}
    (\hat{\textbf p}-q\textbf A)^2\psi=-\hbar^2\nabla^2\psi+2i\hbar q\textbf A\cdot\nabla\psi+i\hbar q\psi\nabla\cdot\textbf A+q^2\textbf A^2\psi
\end{equation*}
since $\nabla\cdot(\psi\textbf A)=\textbf A\cdot\nabla\psi+\psi\nabla\cdot\textbf A$. Now, using Eq. \eqref{eq:mass_vector_potential_connection}, we can replace  $q\nabla\cdot \textbf A$ with $\partial_t \tilde m$ in the equation above. Once this equation is injected into Eq. \eqref{eq:expanded_KG}, we get the Klein-Gordon equation for non-conservative effects
\begin{equation}\label{eq:generalized_KG}
\begin{split}
    i\hbar \tilde m\partial_t\psi+\frac{\hbar^2}2\Box\psi-i\hbar q\textbf A\cdot\nabla\psi=&\\
    \left[\tilde mq\phi+\frac{q^2}{2}\left(\frac{\phi^2}{c^2} + \textbf A^2\right)\right]&\psi,
\end{split}
\end{equation}
where $\Box=\nabla^2-c^{-2}\partial_{tt}$ is the d'Alembertian operator. When only conservative effects are considered, $\tilde m\rightarrow m$, and Eq. \eqref{eq:generalized_KG} turns into the Klein-Gordon equation with relativistic mass.

\section*{Conservation of the probability density}
\subsection*{Conservation equation}
We construct the probability current density using Noether's framework \cite{noether1918,kosmann2011noether} rather than using the usual method \cite{greiner2000relativistic} involving $\psi$ and its complex conjugate inside Eq. \eqref{eq:generalized_KG}, as the mass is not constant. We start with the first order Euler-Lagrange equation given by
\begin{equation}\label{eq:euler_lagrange}
    \frac{\delta\mathcal{L}}{\delta\psi_{i}}-\partial_{\nu}\frac{\delta\mathcal{L}}{\delta\partial_{\nu}\psi_{i}}=0 \,\,\, \forall i\in \{1,\dots,N\},
\end{equation}
where $\mathcal L$ is the Lagrangian density, and we used summation for repeated Greek index $\nu\in\{t,x,y,z\}$. Here $N=2$, since we will use only the function $\psi$ and its complex conjugate $\psi^*$. We can verify that
\begin{equation}\label{eq:lagrangian_density}
\begin{split}
    \mathcal{L}=&-i\frac{\hbar \tilde m}2 (\psi^* \partial_t\psi - \psi\partial_t\psi^*)\\
    &+ \frac{\hbar^2}2 (\nabla\psi^*\cdot\nabla\psi- \frac{1}{c^2} \partial_t\psi \partial_t\psi^*)\\
    &+ i \frac{q\hbar}2 (\psi^*\textbf A\cdot\nabla\psi - \psi\textbf A\cdot\nabla\psi^*)\\
    &+\left[\tilde mq\phi+\frac{q^2}2\left(\frac{\phi^2}{c^2}+\textbf A^2\right)\right]\psi^*\psi,
\end{split}
\end{equation}
used in Eq. \eqref{eq:euler_lagrange} gives Eq.\eqref{eq:generalized_KG} using the Lorenz gauge. With the first order Euler-Lagrange equation, there is a corresponding Noether's current density for $\mu,\nu\in\{t,x,y,z\}$
\begin{equation}\label{eq:Noether_currents}
    j_\mu = \sum_{i}\frac{\delta\mathcal{L}}{\delta\partial_{\mu}\psi_{i}}\delta\psi_{i},
\end{equation}
where $\delta\psi$ and $\delta\psi^*$ correspond to infinitesimal variations under a $U(1)$ phase transformation, i.e. $\psi\rightarrow e^{i\theta}\psi$ and $\psi^*\rightarrow e^{-i\theta}\psi^*$. In this case, $\delta\psi\rightarrow i\theta\psi$, $\delta\psi^*\rightarrow -i\theta\psi^*$, $\partial_{\nu}\delta\psi=i\theta\partial_{\nu}\psi$, and $\partial_{\nu}\delta\psi^*=-i\theta\partial_{\nu}\psi^*$. 
Computing the four currents we get
\begin{equation}\label{eq:four_current_density}
\left.
\begin{split}
    J_t&=\hbar\theta\left(\tilde m\psi\psi^{*}-i\hbar\frac{\psi\partial_{t}\psi^{*}-\psi^{*}\partial_{t}\psi}{2c^2}\right)\\
    J_\mu&=\hbar\theta\left(i\hbar\frac{\psi\partial_\mu\psi^{*}-\psi^{*}\partial_\mu\psi}2-qA_\mu\psi\psi^{*}\right)
\end{split}
\right\}
\end{equation}
with $\mu\in\{x,y,z\}$. From Eq. \eqref{eq:Noether_currents} we can infer that the term inside the square brackets on the RHS of Eq. \eqref{eq:lagrangian_density} does not contribute to the Noether currents. 
Since the Noether theorem guarantees that 
\begin{equation}\label{eq:Noether_conservation_equation}
\partial_{\nu}J_{\nu} =0,
\end{equation}
we can drop $\hbar\theta$ from the four-current density. 
\subsection*{Polar form of the conservation equation}
We can write this equation in polar form using $\psi= R e^{i S/\hbar}$. Here we take the modulus $R(t,x,y,z):\mathbb R^4\rightarrow \mathbb R^+$ and the phase \mbox{$ S(t,x,y,z):\mathbb R^4\rightarrow \mathbb R$}, as this approach is compatible with the quantum operators $\hat{\textbf p}$ and $\hat H$ (e.g. Ref. \citenum{madelung1926}). We can now compute explicitly the current density using the polar form of $\psi$. When $ R\ne0$, we have
\begin{equation*}
    \forall\nu\in\{t,x,y,z\},\,\frac{\partial_\nu\psi}{\psi}=\partial_\nu\ln\psi,
\end{equation*}
where $\ln\psi=\ln R+i S/\hbar$. So, $\psi\partial_\nu\psi^{*}-\psi^{*}\partial_\nu\psi=-2i R^2\partial_\nu S/\hbar$, even when $R=0$. Using this result, the four-current density becomes
\begin{equation}\label{eq:four_current_density_with_S_and_R}
\left.
\begin{split}
    J_t&=\left[\tilde m-\partial_t S/c^2\right] R^2\\
    J_\mu&=\left[\partial_\mu S-qA_\mu\right] R^2
\end{split}
\right\}
\end{equation}
where $\mu\in\{x,y,z\}$.
Using Eqs. \eqref{eq:mass_potential_connection} and \eqref{eq:four_current_density_with_S_and_R}, Eq. \eqref{eq:Noether_conservation_equation} yields
\begin{equation}\label{eq:conservation_with_R_and_S}
    \partial_t \rho+
    \textbf v_\psi\cdot\nabla \rho=-\frac{\rho c^2}{\tilde mc^2-\partial_t S} \Box S.
\end{equation}
where
\begin{equation}\label{eq:particle_velocity}
\textbf v_{\psi}=
    \frac{(\nabla S-q\textbf A)c^2}{\tilde mc^2-\partial_t S},
\end{equation}
and the density $\rho$ is $R^2$. Using Eq. \eqref{eq:mass_vector_potential_connection}, we find
\begin{equation}\label{eq:divergence_of_v}
\begin{split}
    \nabla\cdot\textbf v_\psi=\frac{c^2\nabla^2S-\partial_t\tilde mc^2-\textbf v_\psi\cdot\nabla\left(\tilde mc^2-\partial_tS\right)}{\tilde mc^2-\partial_tS}.
\end{split}
\end{equation}
Adding $\rho\nabla\cdot\textbf v_\psi$ on both sides of Eq. \eqref{eq:conservation_with_R_and_S} yields
\begin{equation}\label{eq:continuity_equation_long_form}
\begin{split}
    \partial_t \rho+
    \nabla\cdot(\rho\textbf v_\psi)=\\-\frac{\partial_t\left(\tilde mc^2-\partial_tS\right)+\textbf v_\psi\cdot\nabla\left(\tilde mc^2-\partial_tS\right)}{\tilde mc^2-\partial_t S}\rho.  
\end{split}
\end{equation}

\subsection*{A necessary and sufficient condition for the conservation of the probability density}
At this point, we can easily identify the momentum of $\psi$ as $\textbf p_\psi=\hbar\textbf k$, and its energy as $E_\psi=\hbar\omega$, where the angular wave vector\cite{goldstein2001classical} is given by \begin{equation}\label{eq:wavevector}
    \textbf k=\nabla S/\hbar,
\end{equation}
and the angular frequency\cite{goldstein2001classical} is
\begin{equation}\label{eq:angular_frequency}
    \omega=-\partial_t S/\hbar.
\end{equation}
So, we can write Eq. \eqref{eq:particle_velocity} as
\begin{equation}\label{eq:new_particle_velocity}
\textbf v_{\psi}=
    \frac{(\textbf p_\psi-q\textbf A)c^2}{\tilde mc^2 +E_\psi}.
\end{equation}
As this equation shows, $\textbf v_\psi$ conforms to the definition of a velocity according to special relativity. Using this insight, we see that the LHS of Eq. \eqref{eq:continuity_equation_long_form} is, in fact, a continuity equation for the density $\rho$ inside a velocity field $\textbf v_\psi$. The RHS of Eq. \eqref{eq:continuity_equation_long_form} is an advection equation (i.e. material derivative) for the energy $\tilde mc^2+ E_
\psi$, and $\textbf v_{\psi}$ is the speed at which this energy is advected. While both equations are conservation equations, there is a fundamental difference. The continuity equation implies conservation inside a static volume while the advection equation implies conservation inside a volume following along the streamlines of the vector field $\textbf v_\psi$.


The necessary and sufficient condition required to conserve the density $\rho$ can be found in Eq. \eqref{eq:continuity_equation_long_form}, namely  
\begin{equation}\label{eq:conservation_of_energy}
\begin{split}
    \partial_t\left(\tilde mc^2+ E_\psi\right)+\textbf v_\psi\cdot\nabla\left(\tilde mc^2+ E_\psi\right)=0.
\end{split}
\end{equation}
Using Eq. \eqref{eq:non_conservative_energy} we find that the equation above is equivalent to,  
\begin{equation}\label{eq:other_conservation_of_energy}
\begin{split}
    \partial_t\left(E_\psi-q\phi\right)+\textbf v_\psi\cdot\nabla\left(E_\psi-q\phi\right)=0.
\end{split}
\end{equation}
So, if the effective wave energy $E_\psi-q\phi$ is conserved then
\begin{equation}\label{eq:conservation_of_density}
    \partial_t \rho+
   \nabla \cdot(\rho \textbf v_\psi)=0.
\end{equation}
Hence the density $\rho$ is conserved in the sense of the continuity equation, and it is always positive when the energy in conserved. As a result, $\rho$ can be interpreted as a probability density, and $\psi$ is a probability amplitude.

\subsection*{Dispersion relation}
When $\Box R=0$, Eq. \eqref{eq:expanded_KG} has a simple  dispersion relation (see the appendix)
\begin{equation}\label{eq:dispersion_relation_with_S}
      \left(\tilde mc^2-\partial_tS\right)^2=m_0^2c^4+(\hbar\textbf k-q\textbf A)^2c^2. 
\end{equation}
In this case, Eq. \eqref{eq:dispersion_relation_with_S} guarantees that $\tilde mc^2-\partial_tS\ne0$ for massive particles. We can use Eqs. \eqref{eq:wavevector} and \eqref{eq:angular_frequency} to write the dispersion relation explicitly as a function of the wave vector and the angular frequency,  
\begin{equation}\label{eq:dispersion_relation}
      \omega_\pm=-\frac{\tilde mc^2}{\hbar}\pm\sqrt{\left(\frac{m_0c^2}{\hbar}\right)^2+ \left(\textbf k-\frac{q}{\hbar}\textbf A\right)^2c^2}.
\end{equation}
This is the de Broglie dispersion relation for matter-wave \cite{debroglie1924recherches}, where the bias term $\tilde mc^2/\hbar$ comes from using $U$ rather than $E$. Under these conditions, we can now compute the group velocity of the probability amplitude of the particle 
\begin{equation}\label{eq:group_velocity_+}
     \frac{\partial\omega_+}{\partial\textbf k}=\frac{(\hbar\textbf k -q\textbf A)c^2}{\sqrt{m_0^2c^4+ \left(\hbar\textbf k-q\textbf A\right)^2c^2}}.
\end{equation}
and
\begin{equation}\label{eq:group_velocity_-}
     \frac{\partial\omega_-}{\partial\textbf k}=\frac{(-\hbar\textbf k+q\textbf A)c^2}{\sqrt{m_0^2c^4+ \left(-\hbar\textbf k+q\textbf A\right)^2c^2}}.
\end{equation}
The positive group velocity is the velocity of the particle with charge $q$ and the negative group velocity is the velocity of the dual particle with charge $-q$. We see here that the particle and its dual particle travel in opposite directions. Note that we are not using the term ``antiparticle" for the dual particle here because we have made no assumption on the actual value of the charge $q$. It can be positive or negative. The only requirement is this paper is that $q\phi<0$. As a result, we see that the particle has an energy $\tilde mc^2+E_\psi>0$ and its dual particle has an energy $\tilde mc^2+E_\psi<0$. Since $q$ must satisfy $q\phi<0$ everywhere, then dual particles cannot propagate in this potential since their charge is $-q$ as it would violate Eq. \eqref{eq:energy_conservation}. As a result, this work only applies to the particle only. 

Using the dispersion relation above we also find that the velocity $\textbf v_\psi$ defined in Eq. \eqref{eq:particle_velocity} is the group velocity given by Eq. \eqref{eq:group_velocity_+} if $\tilde mc^2+E_\psi>0$ and Eq. \eqref{eq:group_velocity_-} if $\tilde mc^2+E_\psi<0$. Note that the amplitude $R$ has a phase velocity $c$ since $\Box R=0$. 

So, the particle is formed by a group of waves and its most probable location is where these waves interfere constructively, leading to a match between the particle velocity and the wave group velocity. Since $\textbf v_\psi$ matches exactly the velocity of a relativistic classical particle in this section, we also conclude that quantum effects are inexistent when $\Box R=0$.

In reality, we know that particles and antiparticles can propagate in electrical potentials of any sign, the restriction herein comes from the fact that we only consider a single particle in this work. So the restriction of the sign of $q\phi$ are stringent. A many-body version of Eq. \eqref{eq:final_generalized_KG} would be required to capture the time evolution of the probability amplitude for a particle and its antiparticle inside a potential of arbitrary sign, a topic beyond the scope of this paper.

\subsection*{Discussion about energy exchange}
In relativity, mass and energy are exchangeable, as a result, we expect Eq. \eqref{eq:energy_conservation} to be violated when mass is converted to energy or vice versa. However, when the wave gains or loses energy from $\tilde m$ (i.e. conservative mass acceleration/deceleration or non-conservative electromagnetic effects, see Eq. \eqref{eq:conservation_of_energy}), or from the potential energy $q\phi$ (i.e. conservative time-independent or non-conservative time-dependent effects, see Eq. \eqref{eq:other_conservation_of_energy}), then the energy is conserved because there is there is really no other energies to tap on here, especially because interactions with other particles (e.g. pair creation/annihilation) cannot be captured by Eq. \eqref{eq:generalized_KG}.




\section*{An interpretation of the space-time coupling}
In this section, we highlight the significance of the second order time derivative inside Eq. \eqref{eq:generalized_KG}. We will assume here that we can divide by the mass operator and only consider conservative processes. we can rewrite this equation as
\begin{equation}\label{eq:final_generalized_KG}
\begin{split}
\partial_t\psi=&\frac{\hbar}{-2im}\Box\psi+\frac{q}{m}\textbf A\cdot\nabla\psi\\
&+\frac{1}{i\hbar}\left[q\phi+\frac{q^2}{2m}\left(\frac{\phi^2}{c^2}+\textbf A^2\right)\right]\psi.
\end{split}
\end{equation}
First, it is interesting to note that time and space are treated on equal footings in Eq. \eqref{eq:final_generalized_KG} when $A\ne0$. Further, if $c\rightarrow\infty$ and $m\rightarrow m_0$, this equation reduces to the Schr\"odinger equation under any conditions. This is a departure of the Klein-Gordon equation with rest mass, which is a relativistic extension of the Schrodinger equation only for a free particle. Since the parallel is quite obvious, the physical meaning of the terms common to both equations will not be discussed here. However, the Klein-Gordon equation with relativistic mass has a second order time derivative, which is not present in the Schr\"odinger equation. We will now use an analogy with electromagnetism to highlight its physical significance.

For this, we look at an electromagnetic field with a scalar potential $\phi_E$ and vector potential $\textbf A_M$. We have used subscripts here to make sure there is no confusion with the potentials $\phi$ and $\textbf A$ used in the previous sections. In the presence of a material with conductivity $\sigma$ and permittivity $\varepsilon$, we can define a ``damped" Lorenz gauge (e.g. see Ref. \citenum{tyler2004three}), derived from the standard Lorenz gauge, as
\begin{equation}\label{eq:damped_Lorenz_gauge}
    \partial_t\phi_E+\mathscr C^2\nabla\cdot\textbf A_M+\frac{\mathscr C^2}{K}\phi_E=0
\end{equation}
Here $\mathscr C$ is the speed of light inside the material, given by $\mathscr C=\sqrt{\mu_0\varepsilon}^{-1}$, and 
$K$ is the magnetic diffusivity given by $K=1/(\mu_0\sigma).$
When the scalar potential is oscillatory, i.e. $\phi_E=\phi_E(\textbf r) e^{-i\omega t}$, Eq. \eqref{eq:damped_Lorenz_gauge} turns into $\nabla\cdot\textbf A_M+\mu_0\sigma^*\phi_E=0$, allowing to define a complex conductivity $\sigma^*=\sigma-i\omega\varepsilon$. Complex conductivities are often used inside the Drude model \cite{drude1900elektronentheorieI,drude1900elektronentheorieII}, where it is interpreted as a lag between the drive and the response \cite{ashcroft1976solid}. In vacuum, where $\sigma=0$ and $\varepsilon=\varepsilon_0$ the magnetic diffusivity, given by
\begin{equation}\label{eq:complex_diffusivity}
    K=\frac{c^2}{-i\omega_V},
\end{equation}
is now purely complex, due to the presence of an oscillating background potential with oscillation frequency $\omega_V$. Under these conditions, the vector potential becomes \cite{tyler2004three}
\begin{equation}\label{eq:wave_equation_for_vector_potential}
    \partial_t\textbf A_M=K\Box\textbf A_M.
\end{equation}
We can now compare the term in front of the $\Box$ operator in Eqs. \eqref{eq:final_generalized_KG} and \eqref{eq:wave_equation_for_vector_potential}. 
We find that
\begin{equation*}
    K=\frac{\hbar}{-2im}
\end{equation*}
in Eq. \eqref{eq:final_generalized_KG}, and the frequency $\omega_V$ defined in Eq. \eqref{eq:complex_diffusivity} is given by
\vspace{-5 pt}
\begin{equation*}
    \omega_V=\frac{2mc^2}{\hbar}.
\end{equation*}
This frequency is exactly the quantum vacuum fluctuation frequency caused by particle and antiparticle pair creation/annihilation both with masses $m$. From the parallel between Eq. \eqref{eq:wave_equation_for_vector_potential} and Eq. \eqref{eq:final_generalized_KG}, we can infer that the complex conductivity $\sigma^*=i\omega_V\varepsilon_0$ creates a delay between the incoming (drive) and outgoing (response) waves in Eq. \eqref{eq:final_generalized_KG} at any given location.  

\section*{Conclusion}
This work derived the Klein-Gordon equation with relativistic mass for a charged particle inside an electromagnetic field. Because the relativistic mass is not Lorentz invariant, this equation cannot describe spinless particles, which require Lorenz invariance. Upon inspection, we found that this equation conserves the probability density and it is always positive, as long as the energy is conserved. In particular, this conservation implies that this equation can be applied to long-lived particles. What is even more interesting, we find that the Schr\"odinger equation is the quasi-static (i.e. $c\rightarrow+\infty$), non-relativistic limit ($m\rightarrow m_0$) version of this equation under any conditions. So, any model built upon the Schr\"odinger equation assumes that the speed of light is infinite, and any results obtained from it must have a superluminal behavior.
In this context, we can say that the Klein-Gordon equation with relativistic mass is a kind of relativistic Schr\"odinger equation. 
Finally, a comparison with electromagnetic theory showed that the second order space-time coupling present in this equation cannot happen without quantum vacuum fluctuations. While not discussed here, the Klein-Gordon equation with relativistic mass can also accommodate spin, using Pauli's approach \cite{pauli1927}, where the quantum operator $(\hat{\textbf p}-q\textbf A)^2$ is replaced by $[\boldsymbol\sigma\cdot(\hat{\textbf p}-q\textbf A)]^2$. 
\section*{Acknowledgments}
This research was supported by the NSF CAREER award PHY-1943939, and the NSF award PHY-2409038.
\section*{appendix}
From Eq. \eqref{eq:non_conservative_energy} we have $ \tilde mq\phi+q^2\phi^2/2c^2=(m_0^2c^4-\tilde m^2c^4)/2c^2$. Using this equation, the polar form of the probability amplitude, and the Lorenz gauge, Eq. \eqref{eq:generalized_KG} yields
\begin{equation*}
\begin{gathered}
\left((\hbar\nabla S-q\textbf A)^2c^2+\left[m_0^2c^4-(\tilde mc^2-\partial_t S)^2\right]\right)R
\\-\hbar^2c^2\Box R =i\hbar c^2 R \Box S\\+2i\hbar \left [(\tilde mc^2 -\partial_t S)\partial_t R+c^2(\nabla S-q\textbf A)\cdot\nabla R\right ], 
\end{gathered}
\end{equation*}
where we dropped $e^{iS/\hbar}$ from both sides of the equation. Since $R$ and $S$ are real, we can split the equation above into
\begin{equation*}\label{eq:other_conservation}
\Box S=\frac{-2}{c^2 R} \left [(\tilde mc^2 -\partial_t S)\partial_t R+c^2(\nabla S-q\textbf A)\cdot\nabla R\right ]
\end{equation*}
and
\begin{equation*}\label{eq:full_dispersion_relation}
\Box R=\left((\nabla S-q\textbf A)^2c^2+m_0^2c^4-(\tilde mc^2-\partial_t S)^2\right)\frac{R}{\hbar^2c^2}.
\end{equation*}
The first equation times $R^2$ is simply the conservation equation Eq. \eqref{eq:conservation_with_R_and_S}.
While, the second equation should be used to compute the time evolution of $R$, it can be seen as the dispersion relation for solutions where $\Box R=0$. 

\bibliography{apssamp}
\end{document}